\documentclass[12pt]{article}

\usepackage{graphicx}

\usepackage{latexsym,array}
\usepackage{amsmath,amssymb}
\usepackage{bm}

\begin{document}
\begin{titlepage}
%
\begin{flushright}
UT-HET 012

KEK-TH-1264
\end{flushright}
\vspace{1\baselineskip}
\vspace*{5mm}
\begin{center}
{\Large \textbf{Light Higgs boson scenario \\
\vspace*{2mm}
in the SUSY seesaw model} }

\vspace{0.5cm} 
{\large
Masaki Asano\,$^{a}$, Takayuki Kubo\,$^{b,c}$, Shigeki Matsumoto\,$^{d}$, \\[0.1cm] 
and Masato Senami\,$^{a,e}$}\\[0.4cm]

{$^{a}$\textit{ICRR, University of Tokyo, Kashiwa, Chiba 277-8582, Japan}}\\[0.3cm]

{$^{b}$\textit {Theory Group, KEK, Tsukuba, Ibaraki 305-0801, Japan}}\\[0.3cm]

{$^{c}$\textit{The Graduate University for Advanced Studies \\[0.1cm]
      (Sokendai), Tsukuba, Ibaraki 305-0801, Japan}}\\[0.3cm]

{$^{d}$\textit{Department of Physics, University of Toyama,
            Toyama 930-8555, Japan}}\\[0.3cm]

{$^{e}$\textit{Department of Micro Engineering, \\[0.1cm]
      Kyoto University, Kyoto 606-8501, Japan}}\\[0pt] 
\vspace*{0.3cm} 

\begin{abstract}
It is demonstrated that the light Higgs boson scenario,
which the lightest Higgs mass is less than the LEP bound, $m_h > 114.4\,$GeV,
is consistent with the SUSY seesaw model.
With the assumptions of the universal right-handed neutrino mass
and the hierarchical mass spectrum of the ordinary neutrinos,
the bounds for the right-handed neutrino mass is investigated
in terms of lepton flavor violating charged lepton decays.
We also discuss the effect of the modification of renormalization group equations
by the right-handed neutrinos on
the $b\to s \gamma$ process and the relic abundance of dark matter
in the light Higgs boson scenario.

\end{abstract}

\end{center}
\end{titlepage}
\newpage 

\section{Introduction}

Supersymmetric (SUSY) models are excellent candidates for
an extension of the standard model (SM) for several reasons.
The quadratic divergent correction to the Higgs mass cancels
between loop diagrams of SM particles and those of SUSY partners.
In SUSY models with R parity conservation,
the lightest SUSY particle (LSP) plays the role of dark matter.
In addition, the grand unification is improved as compared to that in the SM.
Hence, SUSY is well motivated by grand unification theories (GUT).
SUSY models may also explain the difference
of the muon anomalous magnetic moment
between the observed value and the prediction of the SM,
which is 3.4$\,\sigma$ level~\cite{g-2exp,g-2theory}.

Moreover, if right-handed neutrinos are introduced into 
the minimal SUSY SM (MSSM), i.e., the SUSY seesaw model,
the existence and smallness of the ordinary neutrino mass
can be explained naturally~\cite{seesaw}.
In the SUSY seesaw model, many leptogenesis scenarios are discussed
to explain the baryon asymmetry of the universe~\cite{Fukugita:1986hr,leptogenesis,leptogenesis2,D'Ambrosio:2003wy}.
Hence, this model can explain many unsolved problem left in the SM
and is the most celebrated candidate for physics beyond the standard model.

In the MSSM, two Higgs doublet fields ($H_u , H_d$) are required
to provide mass terms for quarks and leptons. 
Hence, three neutral and two charged Higgs bosons exist,
two CP-even scalar ones $h$ (lighter), $H$ (heavier),
one CP-odd pseudoscalar one $A$, and a pair of charged ones $H^\pm$.
If the coupling of $h$ to the $Z$ boson is significantly
smaller than the SM value,
the mass of the lightest one, $m_h$, can be lighter than
the SM bound, $m_h > 114.4 \,\rm GeV$~\cite{LEPSUSY,LHSBottino1,LHSKane}.
This is because the LEP experiments searched primarily
the Higgs boson by $e^+e^- \to Z \to Z h$ mode.
However, the possibility of $m_h < 114.4 \,\rm GeV$, which we call
the light Higgs boson scenario (LHS),  has not been studied thoroughly,
since the LHS cannot be realized in the constrained MSSM (CMSSM).
Studies for the LHS has been restricted in spite of
its potential importance~\cite{LHSBottino1,LHSKane,LHSpheno,Asano:2007gv}.


In our previous work~\cite{Asano:2007gv},
we showed that the LHS scenario
is consistent with the dark matter abundance observed by WMAP~\cite{WMAP}.
In the work, the MSSM with a nonuniversal Higgs mass boundary condition (NUHM) was
adopted as a reference model for the LHS,
which is the simplest one realizing the LHS.
Under this boundary condition,
the masses of Higgs multiplets are different from that of other scalar particles
at the GUT scale, $M_G$.
This is reasonable because the Higgs multiplets are not necessarily
in the same multiplet of a gauge group of GUT with other scalar particles.
The NUHM has six free parameters,
$(m_0, \, m_{1/2}, \, A_0 , \, \tan \beta, \, \mu, \, m_A )$,
where $(m_0, \, m_{1/2}, \, A_0)$ are defined at $M_G$
and the others are defined at the electroweak scale, $m_W$.

Once we introduce the right-handed neutrinos into this model,
i.e. the LHS in the SUSY seesaw model,
lepton flavor violation (LFV) is inevitably induced.
Hence, this model predicts nonzero branching ratios 
for LFV charged lepton decays.
It is known that large $|A_0|$, $\mu$, and $\tan \beta$ enhance
these ratios~\cite{Hisano:1995cp}.
Since the LHS requires large values for these parameters~\cite{Asano:2007gv},
LFV processes are worth investigating,
 when the LHS is discussed in the SUSY seesaw model.
Therefore, we study the effects of the right-handed neutrinos
on the LHS in this work.
In particular, LFV induced by the right-handed neutrinos are studied
in $\mu \to e \gamma, \tau \to \mu \gamma $, and $\tau \to e \gamma $ processes,
and we clarify the constraint for the mass of right-handed neutrinos
with the assumption of the universal right-handed neutrino mass.
Moreover, we show that the LHS parameters obtained in Ref.~\cite{Asano:2007gv}
are not significantly shifted by the modification of
the renormalization group equations (RGE) by the right-handed neutrinos.

This paper is organized as follows.
In the next section, we present the SUSY seesaw model.
We summarize the LFV induced by the modification of
the RGE by the right-handed neutrinos.
In Section~\ref{sec:result},
the LFV processes are discussed in mass insertion approximation.
Then, we show our numerical results for
the lepton flavor violating processes beyond the mass insertion approximation.
In Section~\ref{sec:others}, we investigate the effects of 
the modification of the renormalization group by the right-handed neutrinos
on the allowed region obtained in Ref.~\cite{Asano:2007gv}.
Section~\ref{sec:conclusion} is devoted to summary and discussion.

\section{Model}

In this section we briefly review the SUSY seesaw model.
Then, we mention the modification of the RGE
and LFV induced by the right-handed neutrinos.
We use the notation of the SUSY Les Houches Accord for the MSSM sector~\cite{SLHA1}.

\subsection{Neutrino sector}
The relevant superpotential of the model is given by
\begin{eqnarray}
W = \epsilon_{ab} \bigl[ 
    (Y_E)_{ij} H_1^a L_i^{b} \bar{E}_j  
  - (Y_{\nu})_{ij} H_2^a L_i^{b} \bar{N}_j  
  - \mu H_1^a H_2^b \bigr]
  + \frac{1}{2} M_{ij} \bar{N}_i \bar{N}_j,
\end{eqnarray}
where we take the basis that $Y_E$ and $M$ are diagonalized,
and hence $Y_{\nu}$ is a nondiagonal complex matrix.
In the following, we assume that the Majorana mass of the right-handed neutrino, $M$,
is much larger than the electroweak scale.
After the electroweak symmetry breaking, the neutrino mass matrix is given as
\begin{eqnarray}
\mathcal{L}_{\nu}^{\mathrm{mass}}
= -
\frac{1}{2}
\begin{pmatrix}
\nu_L^T & N^T
\end{pmatrix}
\begin{pmatrix}
0     & m_D \\
m_D^T & M
\end{pmatrix}
\begin{pmatrix}
\nu_L \\
N
\end{pmatrix}
+ \mathrm{h.c.},
\end{eqnarray}
where $ m_D = v_2 Y_{\nu} / \sqrt{2} $ is the neutrino Dirac mass
and $v_2/\sqrt{2}$ is a vacuum expectation value of $H_2$.
Then, the light neutrino mass matrix is given by
\begin{eqnarray}\label{m_nu}
m_{\nu} &\simeq& -m_D M^{-1} m_D^T. 
\end{eqnarray}
This matrix $m_{\nu}$ can be diagonalized by the Maki-Nakagawa-Sakata (MNS) matrix,
\begin{eqnarray}\label{MNS}
\mathrm{diag} (m_{\nu_1},m_{\nu_2},m_{\nu_3})
 = m_{\nu}^{\mathrm{diag}} = U^T_{\mathrm{MNS}} m_{\nu} U_{\mathrm{MNS}}. 
\end{eqnarray}
The matrix $U_{\mathrm{MNS}}$ is defined as
\begin{eqnarray}\label{U=VP}
U_{\mathrm{MNS}} 
&=& V \,\, \mathrm{diag} (e^{-i\frac{\phi}{2}},e^{-i\frac{\phi '}{2}},1 ),
\end{eqnarray}
where $\phi$ and $\phi'$ are CP violating Majorana phases and
$V$ is given by
\begin{eqnarray}
V &=& 
\begin{pmatrix}
c_{12}c_{13}                                & c_{13}s_{12}                                & s_{13}e^{-i\delta} \\
-s_{12}c_{23}-c_{12}s_{23}s_{13}e^{i\delta} & c_{12}c_{23}-s_{12}s_{23}s_{13}e^{i\delta}  & c_{13}s_{23}       \\
s_{12}s_{23}-c_{12}c_{23}s_{13}e^{i\delta}  & -c_{12}s_{23}-c_{23}s_{12}s_{13}e^{i\delta} & c_{13}c_{23}
\end{pmatrix}.
%
\end{eqnarray}
We have abbreviated $\sin \theta_{ij}$ and $\cos \theta_{ij}$
as $s_{ij}$ and $c_{ij}$, respectively. 

Note that from Eqs.~(\ref{m_nu}) and (\ref{MNS}),
$Y_{\nu}$ can be parametrized as follows~\cite{Casas:2001sr},
\begin{eqnarray}\label{Ibarra}
Y_{\nu}^T = i \frac{\sqrt{2}}{v_2} \sqrt{M} R \sqrt{m_{\nu}^{\mathrm{diag}}} U_{\mathrm{MNS}}^{\dagger},
\end{eqnarray}
where $R$ is a $3 \times 3$ complex orthogonal matrix.
For larger right-handed neutrino masses,
larger $Y_\nu$ is derived as $ Y_\nu \propto M^{1/2}$.
As discussed below, the LFV depends crucially on $Y_\nu$.
In the following,
we assume that the right-handed neutrinos have the same mass $M$ for definiteness.
The parameters $m_\nu^{\rm diag}$ and $U_{\rm MNS}$ can be determined by
neutrino oscillation experiments.
We use the following parameters \cite{Strumia, PDG, neutrino},
\begin{eqnarray}
&& m_{\nu_1}=0 \, \mathrm{eV} \, , \nonumber \\
&& \Delta m_{21}^2 = m_2^2 - m_1^2 = 80 \times 10^{-6} \, \mathrm{eV}^2 \, , \nonumber \\
&& |\Delta m_{32}^2| = |m_3^2 - m_2^2| =25 \times 10^{-4} \, \mathrm{eV}^2 \, , \nonumber \\
&& \sin\theta_{12}  =  0.56 \, , \hspace{0.5cm}
   \sin\theta_{23}  =  0.71 \, , \hspace{0.5cm} 
   \sin\theta_{13} \le 0.22 \, ,
\end{eqnarray}
where we assumed the normal hierarchy of the neutrino masses,
$m_{\nu_1} \ll m_{\nu_2} \ll m_{\nu_3}$ and $m_{\nu_1}=0$ for definiteness.
The latter assumption does not affect our results
as long as $ m_{\nu_1} \ll m_{\nu_2}$.
The mixing angle $\theta_{13}$ and the Dirac phase $\delta$
are taken as free parameters.

\subsection{Renormalization group equations}

In addition to the neutrino masses,
the existence of the right-handed neutrinos modifies the RGE of the MSSM.
The RGE between $M_G$ and $M$ are different from those of the MSSM.
By this modification, LFV is induced in the slepton mass matrices
as well as the deviation of masses of superparticles at the electroweak scale
from those without the right-handed neutrinos.

The RGE in the SUSY seesaw model are well known and presented,
for example, in Ref.~\cite{Hisano:1995cp}.
Since the LHS cannot be realized in the CMSSM,
we adopt the boundary condition of the NUHM for solving the RGE.
Except for neutrino parameters,
the NUHM with right-handed neutrinos has seven free parameters,
$(m_0, \, m_{1/2}, \, A_0 , \, \tan \beta, \, \mu, \, m_A , M)$,
where $(m_0, \, m_{1/2}, \, A_0)$ are defined at $M_G$,
$(\tan \beta, \, \mu, \, m_A )$ are defined at $m_W$,
and the universal right-handed neutrino mass $M$
is treated as a constant independent of renormalization scale.
Using the input values at the $m_W$ scale,
a boundary condition at $M_G$ is derived by the renormalization group running.
Then, we solve the RGE with this boundary condition from $M_G$ to $m_W$
and compare the result with the input parameters at $m_W$.
We iterate this procedure
till the result becomes self-consistent with all input parameters.
In this work, the renormalization group running is evaluated by
SPheno~\cite{Porod:2003um}.

In this section, we only mention a LFV part of the RGE. 
LFV is inevitable in this model,
since $Y_E$ and $Y_{\nu}$ cannot be diagonalized simultaneously.
For example, even if $m_{\tilde{L}}^2$ is diagonal at $M_G$,
the LFV is transmitted from neutrino Yukawa couplings
to the soft SUSY breaking mass of the slepton doublet, $m_{\tilde{L}}^2$, through
\begin{eqnarray}\label{SleptonSoftMass}
16\pi^2 \frac{d}{dt} m_{\tilde{L}}^2 
&=& \Bigl[16\pi^2 \frac{d}{dt} m_{\tilde{L}}^2 \Bigr]_{\mathrm{MSSM}}  \nonumber \\
&&  + \left[ 
      (m_{\tilde{L}}^2 Y_{\nu}^* Y_{\nu}^T + Y_{\nu}^* Y_{\nu}^T m_{\tilde{L}}^2 )
      \right. \nonumber \\
&&  \left. +2(Y_{\nu}^* (m_{\tilde{N}}^{2})^T Y_{\nu}^T 
     + m_{H_2}^2 Y_{\nu}^* Y_{\nu}^T + T_{\nu}^* T_{\nu}^T) \right] ,
\end{eqnarray}
where $m_{\tilde N} $ is the soft mass of the right-handed sneutrino,
$ m_{H_2} $ is the mass of the Higgs boson coupling to the up-type quarks,
and $T_{\nu} = A_0 Y_{\nu}$ is the trilinear coupling with the sneutrino.

If all scalar masses at $M_G$ are assumed to be universal
as the CMSSM boundary condition,
the off-diagonal elements of $m_{\tilde{L}}^2$ at low energy are estimated
in the leading log approximation as
\begin{eqnarray}\label{log}
m_{\tilde{L}}^2 \simeq 
\frac{-1}{8\pi^2}(3m_0^2+A_0^2)(Y_{\nu}^* Y_{\nu}^T)\log\frac{M_{\mathrm{GUT}}}{M} 
\label{LFVCMSSM} .
\end{eqnarray}
However, the LHS cannot be realized in the CMSSM,
and therefore we adopted the the boundary condition of the NUHM.
Accordingly, the RGE of this model should be solved numerically.
Nevertheless, it remains true that
$m_{\tilde{L}}^2 $ is almost proportional to $ Y_{\nu}^* Y_{\nu}^T$.

We investigate LFV processes for two parameter sets,
which are summarized in Table \ref{input}.
These parameters are typical ones 
consistent with the observed dark matter abundance
and the branching ratio of $b \to s\gamma$
in the LHS without the  right-handed neutrinos~\cite{Asano:2007gv}. 

\begin{table}[t]
 \begin{center}
  \tabcolsep=3pt \footnotesize
  \begin{tabular}{|c|c|c|c|c|c|c|}
\hline
   Points & $m_0$      & $m_{1/2}$ & $A_0$     & tan$\beta$ & $m_A$   & $\mu$      \\ \hline
   1      & 196.25 GeV & 323 GeV   & $-$1000 GeV & 10         & 104 GeV & 600 GeV    \\ \hline
   2      & 492 GeV    & 147.5 GeV & $-$1000 GeV & 10         & 104 GeV & 600 GeV    \\ \hline   
  \end{tabular}
 \end{center} \vspace{-3mm}
 \caption[123]{\small Input parameters for the LHS}
 \label{input}
\end{table}%
%

\section{Lepton Flavor Violation}
\label{sec:result}

In this section, we study LFV processes in the LHS with the right-handed neutrinos.
In particular, the branching ratios of the LFV processes, $l_i \to l_j \gamma$,
are discussed.

\subsection{Mass insertion approximation}

The LFV in $m_{\tilde{L}}^2$ induces
LFV charged lepton decays, for example, $l_i \to l_j \gamma$ decays.
In the mass insertion approximation,
the contribution to $l_i \to l_j \gamma$ decay amplitude is proportional to 
$(m_{\tilde{L}}^2)_{ij}$. 
As a result, the branching ratios Br$(l_i \to l_j \gamma)$ are proportional to
$|(m_{\tilde{L}}^2)_{ij}|^2$ and roughly estimated as
\begin{eqnarray}\label{massinsertion}
\mathrm{Br}(l_i \to l_j \gamma) 
\propto
\frac{\alpha^3}{G_F^2} 
\frac{|(m_{\tilde{L}}^2)_{ij}|^2}{ m_S^8} \tan^2 \beta
\propto
\frac{\alpha^3 \tan^2 \beta}{G_F^2 m_S^8}
|(Y_{\nu}^* Y_{\nu}^T)_{ij}|^2 , 
\end{eqnarray}
where $ m_S$ means a typical scale of SUSY particle masses,
$\alpha$ is the fine structure constant, and $G_F$ is the Fermi constant. 
In the following, we assume $R$ is a real orthogonal matrix for simplicity.
Then $Y_{\nu}^* Y_{\nu}^T$ is independent of the form of $R$ under 
our assumption of the universal right-handed neutrino mass and is written as
\begin{eqnarray}\label{YY}
Y_{\nu}^* Y_{\nu}^T = 
( {2M} / {v_2^2} ) V m_{\nu}^{\mathrm{diag}} V^{\dagger}.
\end{eqnarray}
Here, we find $Y_{\nu}^* Y_{\nu}^T$ is also independent of Majorana phases.
As a result, it can be seen that the structure of the LFV
is determined by the ordinary neutrino masses and the mixings,
and the right-handed neutrino mass determines the normalization of the LFV. 
Note here that even if we treat $R$ as complex, the branching ratios of the LFV processes
are expected to be in the same order as the real case except for an accidental cancellation due to the phases in $R$.
 

The amplitude of $ \mu \to e \gamma $ is known to be enhanced
for large $A$ and $\mu \tan \beta $ as $ A/ m_S$ and $\mu \tan \beta /m_S$.
These are much larger than $m_S$ in the LHS~\cite{Asano:2007gv},
since $A$ and $\mu$ are larger than $m_0$ and $m_{1/2}$.
In addition, the LHS requires large $\tan \beta$, which also enhances the amplitude.
Moreover, masses of SUSY particles in the LHS
are small, so that $ m_S \sim $ a few hundreds GeV.
Accordingly, the constraint on $M$ in the LHS seems to be severe.

\subsection{Beyond the mass insertion approximation}

In this subsection, we will present our numerical results for
the branching ratios of LFV processes, $l_i \to l_j \gamma$.

\begin{figure}[p]
 \begin{minipage}{0.5\linewidth}
  \centering
  \includegraphics[width=7.7cm,clip]{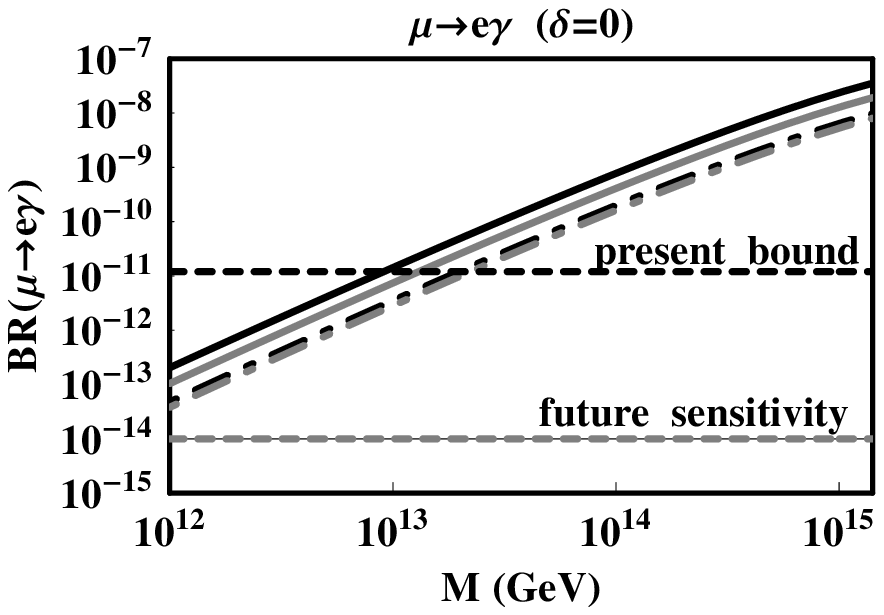}\\
  \includegraphics[width=7.7cm,clip]{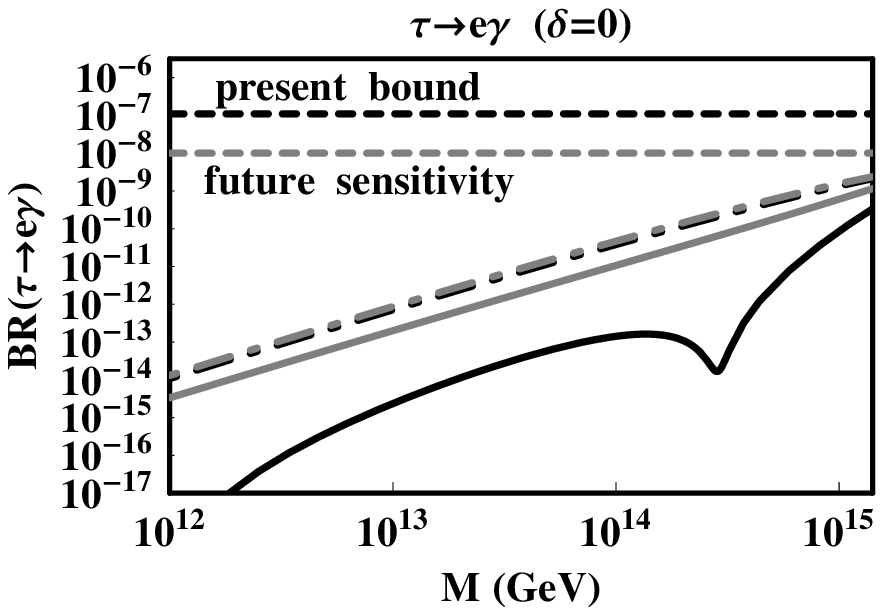}\\
  \includegraphics[width=7.7cm,clip]{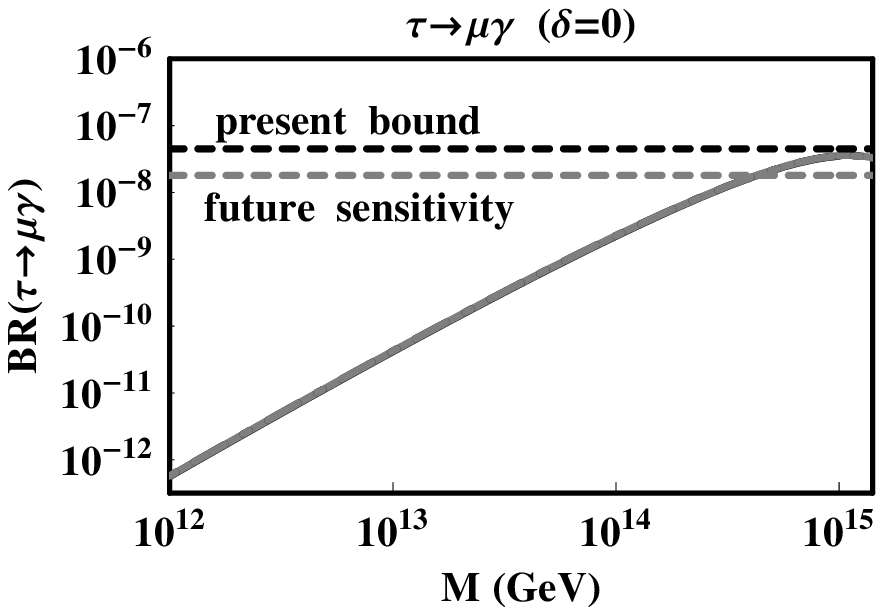}\\
 \end{minipage}
 \begin{minipage}{0.5\linewidth}
  \centering
  \includegraphics[width=7.7cm,clip]{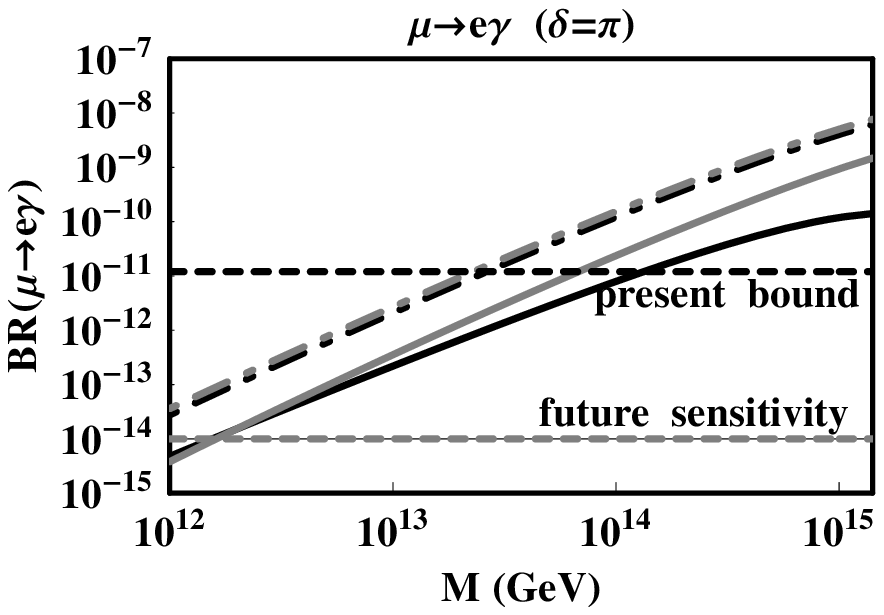}\\
  \includegraphics[width=7.7cm,clip]{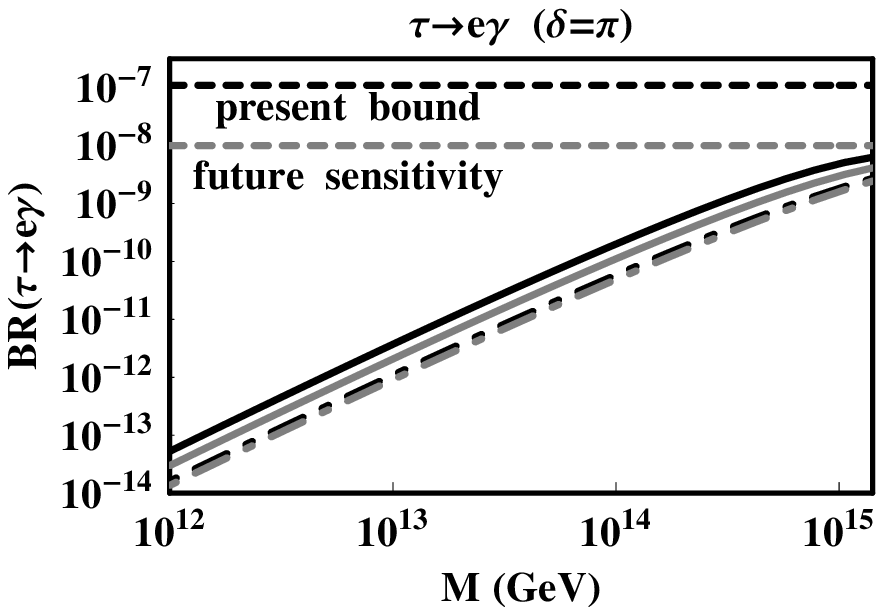}\\
  \includegraphics[width=7.7cm,clip]{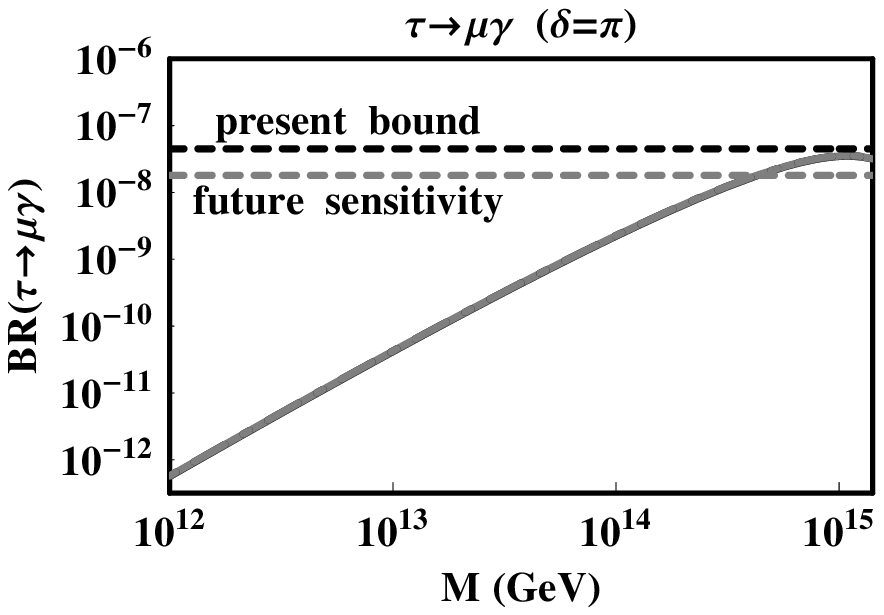}\\
 \end{minipage}
 \caption{\small The branching ratios Br$(l_i \to l_j \gamma)$ at Point 1
 are shown as a function of $M$.
 The solid (dashed-dotted) black and gray lines are for 
 $\sin\theta_{13}=0.1 ~ (0.01)$ and $\sin\theta_{13}=0.05 ~(0.001)$,  respectively.
 The horizontal dashed black and gray lines represent
 the present bound and future sensitivity. }    
 \label{fig:LFV1} 
\end{figure}

With the input parameters shown in Table \ref{input},
we calculate the branching ratios of
$\mu \to e \gamma$, $ \tau \to e \gamma$, and $\tau \to \mu \gamma$ processes.
We calculate these processes beyond the mass insertion approximation
following Ref.~\cite{Hisano:1995cp}.
Figure~\ref{fig:LFV1} shows the branching ratios at Point 1
as a function of $M$.
The left (right) figures are results for $\delta = 0 \ (\pi)$.
The solid (dashed-dotted) black and gray lines are 
for $\sin\theta_{13}=0.1 \ (0.01)$ and $\sin\theta_{13}=0.05 \ (0.001)$, respectively.
The branching ratios of $\mu \to e \gamma$ and $\tau \to e \gamma$ 
depend on $\theta_{13}$ and $\delta$,
while $\tau \to \mu \gamma$ is independent of them.
The horizontal dotted black and gray lines show
the present bound and future sensitivity.
The present bounds are summarized as 
$\mathrm{Br}(\mu  \to e \gamma) < 1.2 \times 10^{-11}$~\cite{muegamma},
$ \mathrm{Br}(\tau \to e \gamma) < 1.1 \times 10^{-7}$~\cite{Belle,tauegamma}, and
$ \mathrm{Br}(\tau \to \mu \gamma) < 4.5 \times 10^{-8}$~\cite{Belle,taumugamma}.\footnote{
More stringent bounds are reported in Ref.~\cite{Banerjee:2007rj}
as $ \mathrm{Br}(\tau \to e \gamma) < 9.4 \times 10^{-8}$ and
$ \mathrm{Br}(\tau \to \mu \gamma) < 1.6 \times 10^{-8}$,
which are combined bounds of the BaBar and Belle results.
The most stringent bound comes still from the $ \mu  \to e \gamma $ process,
even if these bounds are used.
} 
As the future sensitivity, we take 
$\mathrm{Br}(\mu \to e \gamma) \sim 10^{-14}$~\cite{MEG},
and $\mathrm{Br}(\tau \to e \gamma) \sim \mathrm{Br}(\tau \to \mu \gamma)
\sim 10^{-8}$~\cite{Akeroyd:2004mj}.
The most stringent constraint is given by the $\mu \to e \gamma$ process,
whose bound is given as $ M \lesssim 10^{13}$\,GeV. 
This bound can be relaxed as $M \lesssim 10^{14}$\,GeV
in a specific case with large $ \sin\theta_{13} $ and $\delta \sim \pi$.
The behavior of this cancellation is determined by $Y_\nu^* Y_\nu^T $, i.e.,
neutrino masses and mixings.
As a result, Br($\mu \to e \gamma$) takes the smallest value
for $\delta = \pi$ and $\sin \theta_{13} \sim 0.1$,
which are illustrated later.

\begin{figure}[t]
 \begin{minipage}{0.5\linewidth}
  \centering
  \includegraphics[width=7.1cm,clip]{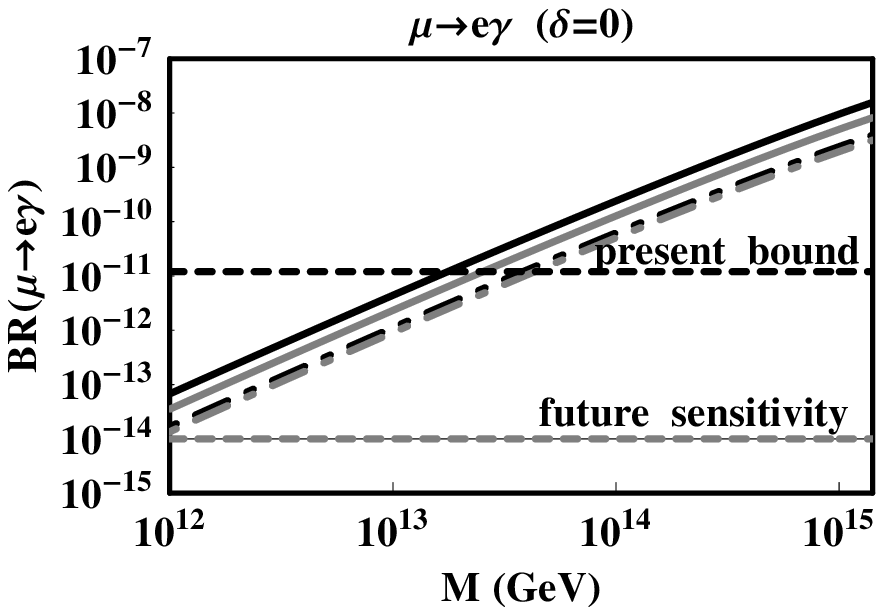}\\
 \end{minipage}
 \begin{minipage}{0.5\linewidth}
  \centering
  \includegraphics[width=7.1cm,clip]{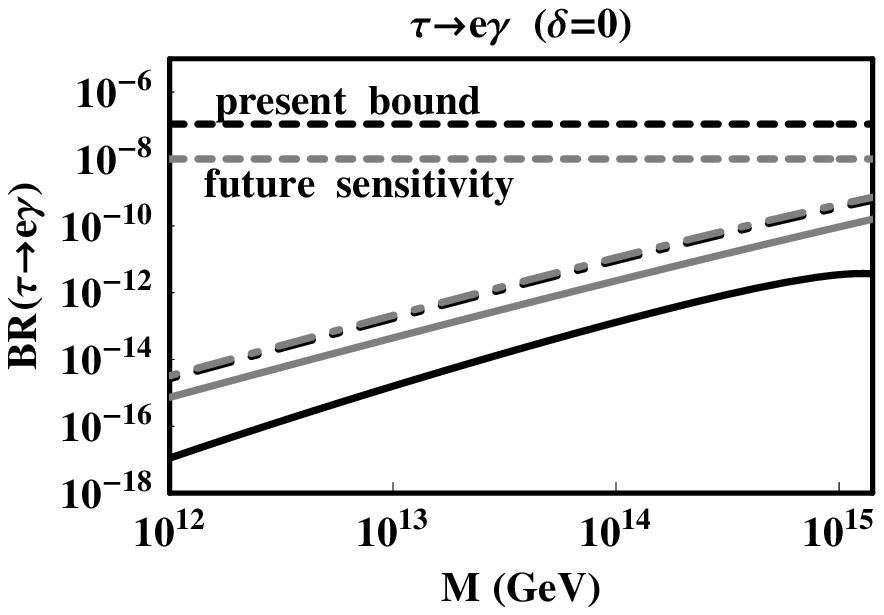}
 \end{minipage}
 \begin{minipage}{0.5\linewidth}
  \centering
  \includegraphics[width=7.1cm,clip]{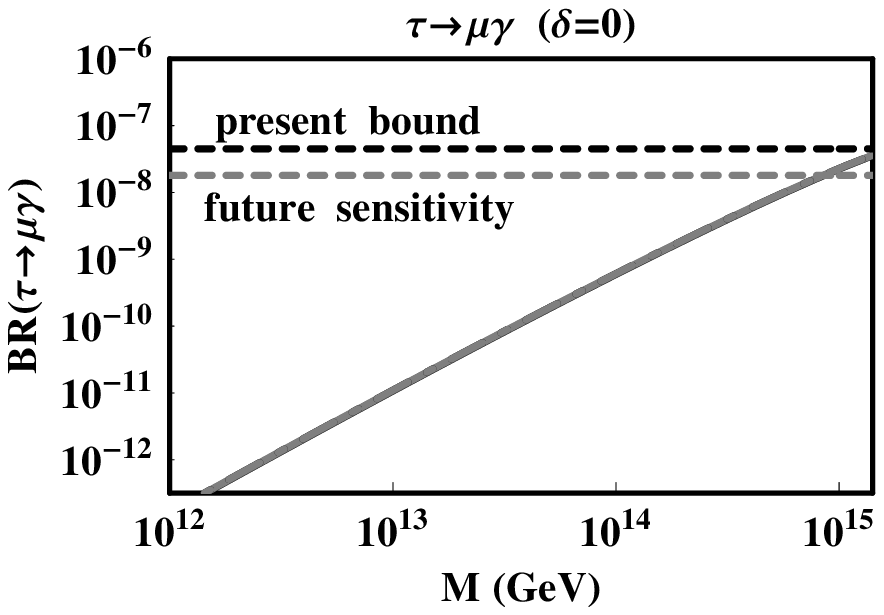}
 \end{minipage}
 \caption{\small The branching ratios Br$(l_i \to l_j \gamma)$ at Point 2
 are shown as a function of $M$.
 The solid (dashed-dotted) black and gray lines are for
 $\sin\theta_{13}=0.1 \, (0.01)$ and $\sin\theta_{13}=0.05 \, (0.001)$, respectively.
 The horizontal dashed black and gray lines represent
 the present bound and future sensitivity. }    
 \label{fig:LFV2} 
\end{figure}

Figure~\ref{fig:LFV2} shows the branching ratios at Point 2.
The dependences of the ratios on $\theta_{13}$ and $\delta$
are almost the same as Point 1,
since its comes almost from $Y_\nu^* Y_\nu^T $.
Hence, we show the results for only $\delta = 0$.
The constraints on $M$ at Point 2 is slightly weak compared to those at Point 1.
This is because the slepton masses at Point 2 are larger than those at Point 1.

\begin{figure}[]
  \centering
  \includegraphics[width=9cm,clip]{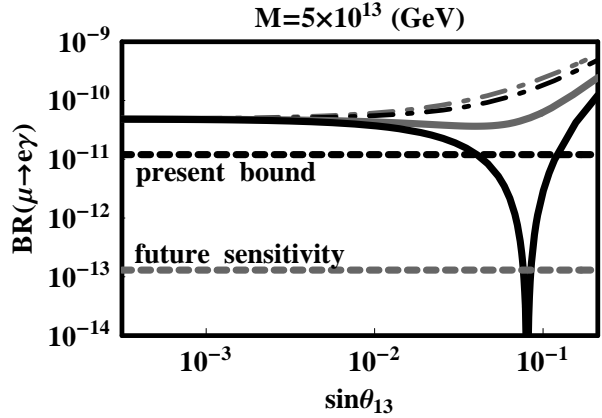}
 \caption{\small The branching ratio of Br$(\mu \to e \gamma)$ at Point 1
 is shown as a function of $\sin \theta_{13}$.
 The solid (dashed-dotted) black and gray lines are
 for $\delta=\pi \, (\pi /3)$ and $\delta=2\pi /3 \, (0)$, respectively.
 The horizontal dotted line represents the present bound and future sensitivity. }    
 \label{fig:BR_s13} 
\end{figure}

In Fig.~\ref{fig:BR_s13}, the branching ratio of the $\mu \to e\gamma$ process
is shown as a function of $\sin \theta_{13}$ at Point 1.
We calculate the branching ratio for several values of $\delta$.
The solid black and gray lines are for $\delta = \pi$ and $\delta=2\pi /3$,
and the dashed-dotted black and gray lines are for $\delta=\pi /3$ and $\delta=0$,
respectively.
The horizontal dashed black and gray lines represent
the present bound and future sensitivity.
The right-handed neutrino mass is taken to be $M = 5 \times 10^{13}$\,GeV.
The results for $\delta > \pi $ is almost the same as $2 \pi - \delta$.
It can be seen that the branching ratio is significantly small
for $\delta \simeq \pi$ and $\sin \theta_{13} \sim 0.1$.
When $\delta \lesssim 2 \pi / 3 $,
the branching ratio is larger for larger $\sin \theta_{13}$.
When $ \delta > 2 \pi / 3 $,
the branching ratio is suppressed for large $\sin \theta_{13}$.
This cancellation is maximum at $\sin \theta_{13} \simeq 0.1$.

%

\section{Other constraints}
\label{sec:others}

Introducing the right-handed neutrinos into the LHS
may shift the allowed parameter region obtained in Ref.~\cite{Asano:2007gv}
due to the modification of the RGE.
In this section, we discuss the modification of the RGE
and show that the allowed parameter region is almost not altered
if $ M $ is constrained by the $ \mu \to e \gamma $ process.

\begin{figure}[]
 \begin{minipage}{0.5\hsize}
  \centering
  \includegraphics[width=7.2cm,clip]{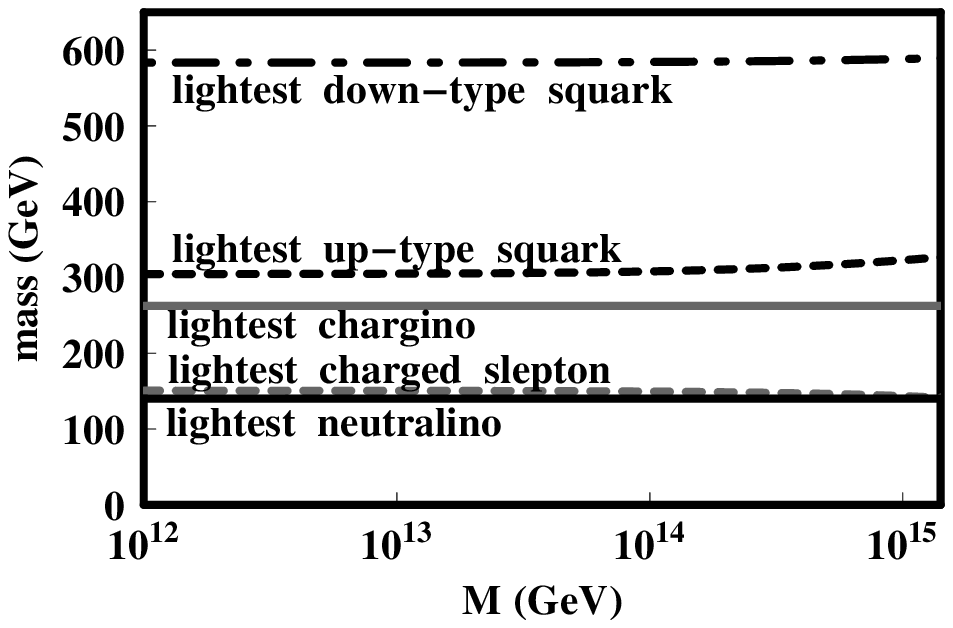}\\
  (a) point 1
 \end{minipage}
 \begin{minipage}{0.5\hsize}
  \centering
  \includegraphics[width=7.2cm,clip]{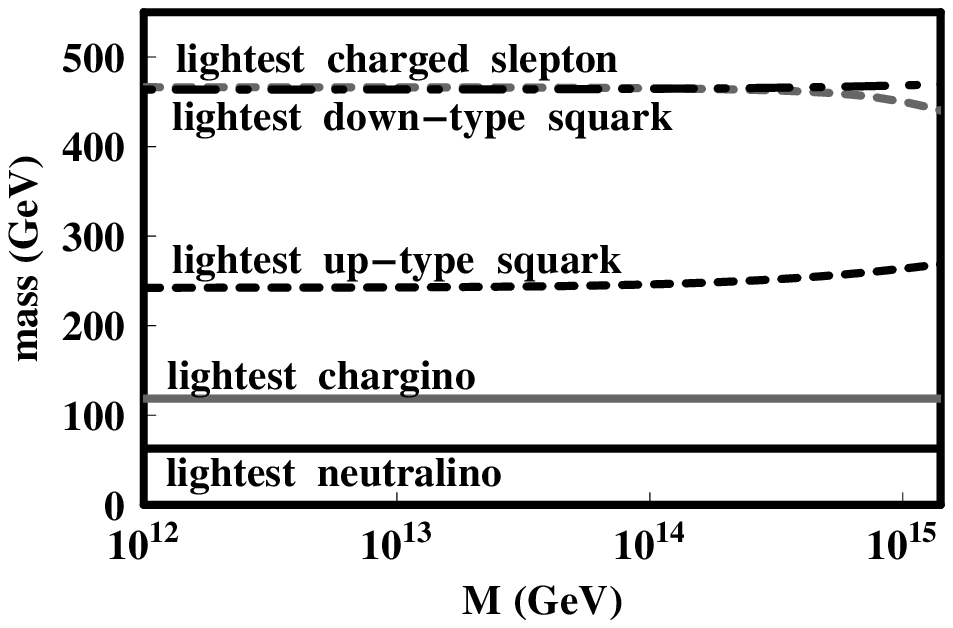}\\
  (b) point 2
 \end{minipage} 
 \caption{\small The masses of the lightest up-type squark,
 the lightest down-type squark, the lightest charged slepton, the lightest neutralino,
 and the lightest chargino are shown as a function of $M$.}    
 \label{fig:masses} 
\end{figure}

The allowed region in the LHS is determined by
the $b\to s \gamma$ process and the relic abundance of dark matter.
First, we discuss the $b\to s \gamma$ process.
The flavor mixing in the squark sector
is not induced by the right-handed neutrinos in the leading order.
Hence, the allowed region of ${\rm Br} ( b \to s \gamma )$
is not shifted if the mass spectrum of the LHS in the SUSY seesaw model
is not changed compared to the case without the right-handed neutrino.
In Fig.~\ref{fig:masses}, the masses of the lightest up-type squark,
the lightest down-type squark, the lightest charged slepton, the lightest neutralino,
and the lightest chargino are shown as a function of $M$.
The left (right) figure is the result at Point 1 (2).
As seen in these figures,
the masses of SUSY particles are not significantly
altered from those without the right-handed neutrino,
since the effect of the right-handed neutrino become negligible for
smaller $M$.
Hence, the parameter region allowed by ${\rm Br} ( b \to s \gamma )$
safely remains unchanged.
In fact, the difference by including the right-handed neutrinos of $m=10^{14}$\,GeV
is only 2 \%. 
Though we have showed masses only for some particles in these figures,
we confirmed that masses of other SUSY particles are also not changed.

\begin{figure}[]
  \centering
  \includegraphics[width=6.7cm,clip]{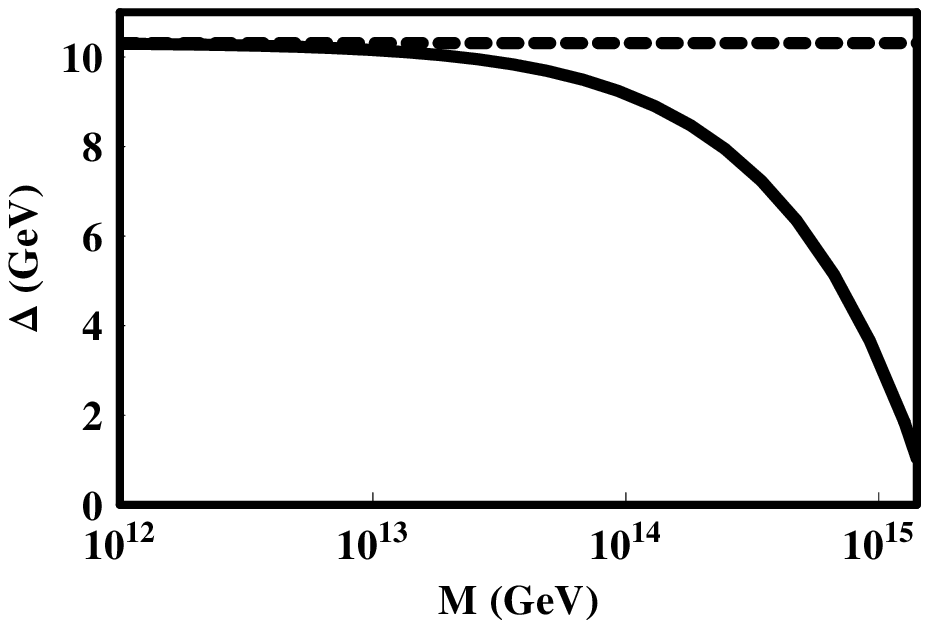}
\includegraphics[width=6.7cm,clip]{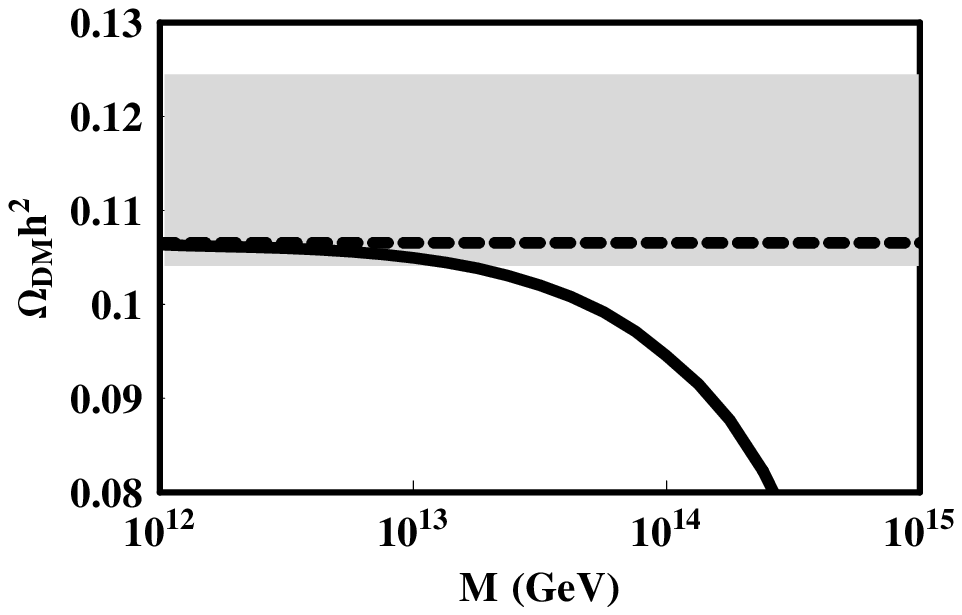}\\
 \caption{\small (Left figure) The mass difference between the lightest neutralino
 and the lightest slepton is shown.
 The mass difference of the LHS with (without) the right-handed neutrinos
 is shown as the solid (dashed) line.
 (Right figure) The relic abundance of the LSP is shown
 as a function of the right-handed neutrino mass.
 The abundance with (without) the right-handed neutrinos
 is solid (dashed) line.
 The shaded region is the region allowed by the WMAP observation. }  
 \label{fig:MassSplitting} 
\end{figure}
Next, we discuss the relic abundance of dark matter~\cite{micrOMEGAs}.
At Point 2, the abundance of the lightest neutralino is
determined by the annihilation process mediated by the pseudoscalar Higgs.
Since the mass of the pseudoscalar Higgs is one of the input parameters,
the abundance at Point 2 is not affected by the existence of the right-handed neutrinos.
On the other hand, Point 1 is in the coannihilation region,
where the relic abundance depends strongly on
the mass difference between the lightest neutralino and the lightest
slepton. 
Note that the mass difference is more sensitive to $M$ than mass itself.
In Fig.\,\ref{fig:MassSplitting} (left), we show the mass difference between
the lightest neutralino $({\tilde \chi}^1_0)$
and the lightest slepton $({\tilde e}_1)$,
$\Delta \equiv m_{\tilde{e}_1} - m_{\tilde{\chi}^0_1}$ as a function of $M$.
The solid (dashed) line shows
$\Delta$ of the LHS with (without) the right-handed neutrinos.
In most of the region allowed by the $\mu \to e \gamma$ process,
the shift of the mass difference is negligible.
At $M\sim10^{14}$\,GeV, the mass difference decreases by a few GeV.
Figure~\ref{fig:MassSplitting} (right) shows the relic abundance of dark matter 
as a function of the right-handed neutrino mass.
The abundance with (without) the existence of the right-handed neutrinos
is denoted by the solid (dashed) line.
The shaded region is the region allowed by the WMAP observation at 3\;$\sigma$ level.
As seen in this figure, the effect of the right-handed neutrinos
on the abundance in the coannihilation region of the LHS
is negligibly small for $M \lesssim 10^{13}$ GeV. 
For $M \sim 10^{14}$\,GeV, which is the very restricted case as above, 
the mass difference deviates significantly from that without the right-handed neutrinos.
Hence, the abundance is significantly smaller than
that without the right-handed neutrinos.
Nevertheless, this is not a severe problem.
The mass of the slepton is dependent on $m_0$, while that of the neutralino is not.
Therefore the increase of $m_0$ by several GeV compensate for the decrease of 
the mass difference by the existence of the right-handed neutrinos.
For this modification of $m_0$, Br$(b \to s \gamma)$ is not changed significantly.
Therefore, after introducing the right-handed neutrinos the allowed region of the LHS 
can survives with slightly larger $m_0$.

We also mention other constraints for the LHS,
$B_s \to \mu^+ \mu^-$ and the muon anomalous magnetic moment.
The deviation due to the existence of the right-handed neutrinos
is small because masses of SUSY particles are not significantly changed.
For the same reason, the prediction for direct detection of dark matter
is not altered.
 
\section{Summary and discussion}
\label{sec:conclusion}
We have demonstrated that the LHS can be realized in the SUSY seesaw model.
With the assumptions of the universal right-handed neutrino mass
and the hierarchical mass spectrum of the ordinary neutrinos,
we have shown that
LFV processes are consistent with the present experimental bounds
if the right-handed neutrino masses are less than $10^{13}$\,GeV.
If $\sin \theta_{13} \simeq 0.1$ and $\delta \simeq \pi$,
this bound is relaxed as $10^{14}$\,GeV.
We have also confirmed that
the relic abundance of the lightest neutralino is not
affected by the introduction of the right-handed neutrinos,
if the right-handed neutrino mass satisfies the LFV bound.

We assumed that the ordinary neutrino mass spectrum is  hierarchical
and the right-handed neutrino mass is universal.
Even if the inverted hierarchical mass spectrum is assumed,
the order of magnitude of  the constraint for the right-handed neutrino mass
is not changed significantly.
While the ratios between ${\rm Br} (\mu \to e \gamma)$,
${\rm Br} (\tau \to e \gamma)$, and ${\rm Br} (\tau \to \mu \gamma)$ are varied,
the $\mu \to e \gamma $ process gives the most stringent constraint
for the right-handed neutrino mass.
On the other hand, if the universal right-handed neutrino mass is not assumed,
the matrix $R$ cannot be ignored even if $R$ is a real matrix.
As a result, the predicting the branching ratios becomes complicated.
In a particular case,
accidental cancellation may occur in a LFV process.
Nevertheless, much severer constraint for the right-handed neutrino mass
is not plausible.

Now, it has been confirmed that the LHS is consistent with
the extension including the neutrino masses.
One of the most important problems left in the SM is
the origin of the baryon number asymmetry of the universe.
In the SUSY seesaw model, the thermal leptogenesis suffers from
the severe gravitino problem~\cite{leptogenesis,gravitino}.
However, in SUSY models, 
the asymmetry can be generated by nonthermal leptogenesis,
for example, the Affleck-Dine mechanism~\cite{leptogenesis2,AD}.
Therefore, the LHS can be consistent with
the baryon number asymmetry of the universe.

\section*{Acknowledgments}

T.K. thank Y.~Okada, T.~Goto, and D.~Nomura for valuable advices.


\end{document}